\documentclass[aps,prl,twocolumn,english,showpacs,10pt]{revtex4-1}

\usepackage[T1]{fontenc}
\usepackage[latin1]{inputenc}
\usepackage{lmodern}
\usepackage{amsmath}
\usepackage{amssymb}
\usepackage{pst-all}
\usepackage{psfrag}
\usepackage{graphicx}
\usepackage{dcolumn}
\usepackage{bm}
\usepackage{epsfig}
\usepackage{braket}
\usepackage{floatrow}

\makeatletter

\begin{document}

\title{Surface plasmon launching by polariton superradiance}
\author{Yu-Xiang Zhang}
\email{iyxz@phys.au.dk}
\author{Yuan Zhang}
\email{yzhang@phys.au.dk}
\author{Klaus M{\o}lmer}
\email{moelmer@phys.au.dk}
\affiliation{Department of Physics and Astronomy, Aarhus University, DK-8000 Aarhus C, Denmark}

\keywords{directional, timed-Dicke states, collective light-matter interaction, 
superradiance, graphene near-infrared plasmon, Non-Markovian dynamics}

\begin{abstract}
The condition of phase matching prohibits the transfer of excitation from free-space photons to
surface plasmon polaritons (SPP). We propose and analyze a scheme that excites an ensemble of emitters
in a collective state, which is phase matched with the SPP by the optical pulses used for its preparation.
By a collective enhancement the ensemble, hence, emits an SPP in a well defined direction.
We demonstrate the scheme by analyzing the launching of near-infrared graphene SPP.
Our theory incorporates the dispersive and dissipative
properties of the plasmon modes to evaluate the non-Markovian emission by the ensembles and will also be
applicable for other types of surface polaritons.
\end{abstract}

\maketitle

Surface plasmon polaritons (SPPs) are electromagnetic modes confined
at metal-dielectric interfaces \cite{Novotny2012} or near two-dimensional materials such
as graphene \cite{Grigorenko2012}. SPPs have dispersion relations different from the ones of free space photons.
This difference, occurring also for phonon polaritons \cite{Hillenbrand2002,Taubner2006,Dai2014,Li2018}, exciton
polaritons \cite{Low2016,Basov2016} and surface polaritons in heterostructures \cite{Lin2017,Woessner2014},
leads to a wavenumber mismatch and
prevents their effective production by conversion from free-space photons.
To close the mismatch and launch SPPs, conventional methods use prisms within the Otto configuration \cite{Otto1968} or the
Kretschmann configuration \cite{Kretschmann1971} to shorten the
photon wavelength, or they equip the SPP dispersion 
relation with band structure by using grating couplers \cite{Raether1988}, or lengthen the SPP wavelength with
an atomic gas medium \cite{Du2015}.
The excitation of graphene SPP is more challenging because
the wavelength of graphene SPP (THz to near-infrared regimes) is two orders of 
magnitude smaller than that of free-space light of the same frequency \cite{Koppens2011}.
Special techniques use scattering resonances of nanoantennas \cite{Krasnok2018,Alonso-Gonzalez2014,Gao_2012,Gao_2013} 
or near-field sources \cite{Fei_2012,Chen_2012}, 
and optical methods, based on the intrinsic nonlinear interaction of graphene with light \cite{Constant_2015},
have realized launching of graphene SPPs in THz to mid-infrared regimes.

In this Letter, we will investigate the prospects of SPP launching by an emitter ensemble.
A single localized two-level quantum emitter may, indeed, absorb an optical photon and subsequently emit
an SPP by spontaneous emission \cite{Akimov2007,Tame2013,Tielrooij_2015}.  
For a point source there is no issue of wavenumber mismatch, but also no control of the directionality of 
the launched SPP. Excitation of SPP with a single wavenumber and direction is
vital for many applications
\cite{Lopez-Tejeira2007,Lin2013,Pors2014,You2015,Krasnok2018,Bliokh2018,Song2017,Andolina2018}.
Our proposal applies a train of $\pi$-pulses to write a wave vector 
into the phase of the spin wave excitation of the emitter ensemble, which is phase matched with the
SPP with the desired directionality (determined by the wave vectors of the $\pi$-pulses).
While we will demonstrate the scheme for a near-infrared graphene SPP, the theory works for a broad range of SPPs and
will be applicable also to other surface polaritons \cite{Hillenbrand2002,Taubner2006,Dai2014,Li2018,Low2016,Basov2016,Lin2017,Woessner2014}.

\begin{figure}[b]
    \includegraphics[width=0.9\textwidth]{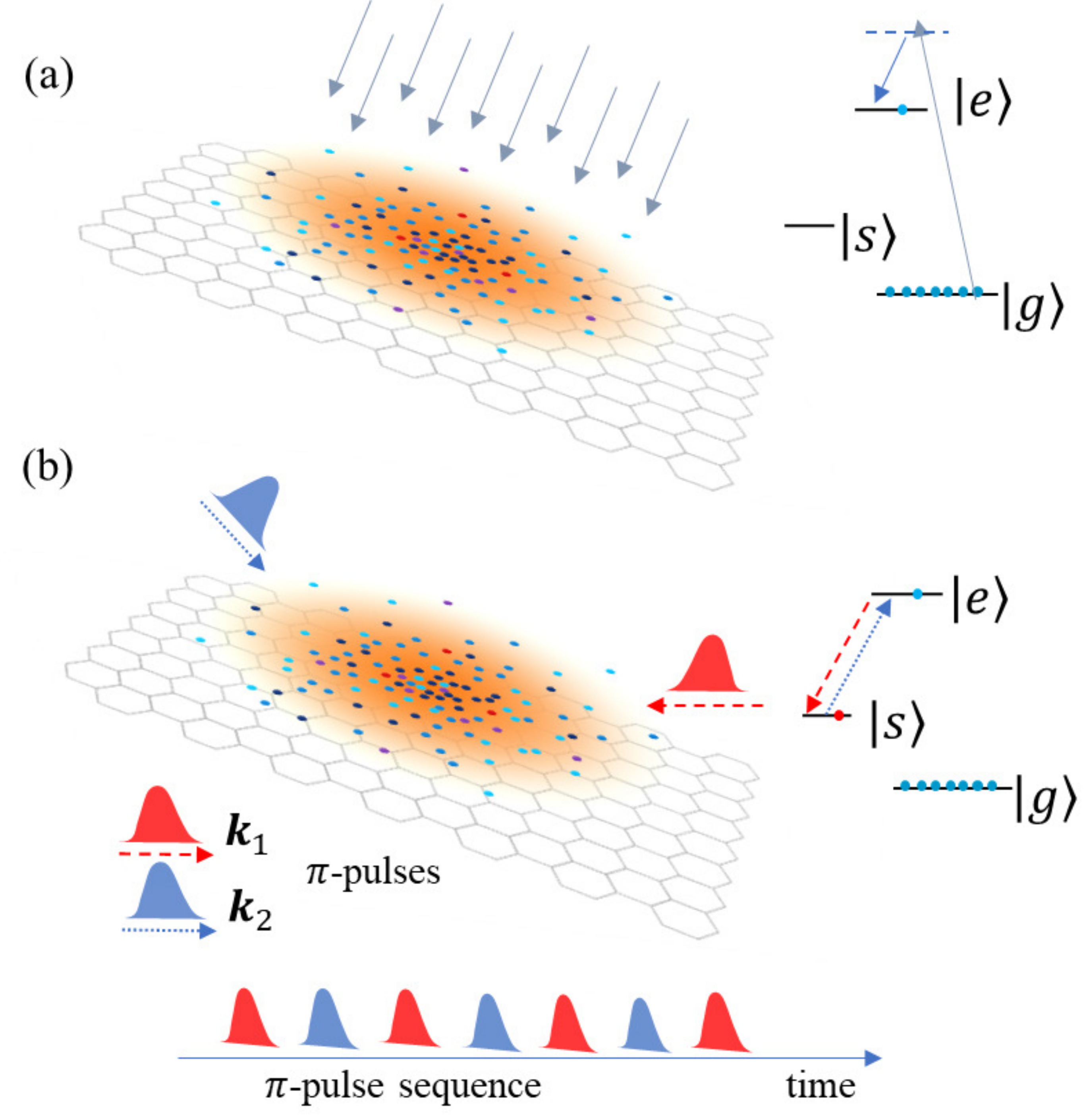}
    \caption{Scheme for preparation of an SPP phase matched timed-Dicke state. (a) Raman excitation of one of the emitters
    into the excited state $|e\rangle$, using uniform optical illumination perpendicular to the surface.
    (b) Illumination by a train of $\pi$-pulses on the $\ket{s}\leftrightarrow \ket{e}$ transition, driving the emitters to
    the target timed-Dicke state $|\psi_{\mathbf{k}_{sp}}\rangle$.}
    \label{fig1}
\end{figure}

When a photon is absorbed by a single emitter, the information of its wave vector
is lost and has no impact on following emission processes.
However, if the photon is uniformly absorbed
by an ensemble of $N$ emitters, its wave vector $\mathbf{k}$ is recorded by the emitters in the phases of the so-called \emph{timed-Dicke state} \cite{Scully2006},
\begin{equation}\label{timed-Dicke}
  |\psi_{\mathbf{k}}\rangle= \frac{1}{\sqrt{N}}
  \sum_{a=1}^{N}\;e^{i\mathbf{k}\cdot\mathbf{r}_a} |s_a\rangle\bigotimes_{b\neq a} |g_{b}\rangle,
\end{equation}
where $|g\rangle$ is the emitter ground state, $|s\rangle$ is an excited state, and
$\mathbf{r}_a$ is the position of the $a^{th}$ emitter.
In analogy with single-photon superradiance \cite{Scully2009} if $|s\rangle$ couples to the SPP field, 
spontaneous emission from $|\psi_{\mathbf{k}}\rangle$ generates a polariton excitation
with wave vector $\simeq \hbar\mathbf{k}$ and energy $\simeq \hbar\omega_{sg}=E_s-E_g$
\cite{Scully2006}. Directional SPP launching based on
this process is possible only if $|\psi_{\mathbf{k}_{sp}}\rangle$ is prepared
with the appropriate SPP wave vector $\mathbf{k}_{sp}$, such that the SPP frequency $\omega_{\mathbf{k}_{sp}}=\omega_{sg}$.
This can be accomplished by Raman processes via  a third atomic level $\ket{e}$, see Fig. \ref{fig1}.

\emph{Preparation of $|\psi_{\mathbf{k}_{sp}}\rangle$}-We consider the simple case where SPPs are confined to an infinite plane interface,
above which a parallel thin layer of emitters is deposited. The preparation of $|\psi_{\mathbf{k}_{sp}}\rangle$ proceeds by two steps.
\begin{enumerate}
\item A single quantum is uniformly absorbed by a Raman process  $|g\rangle \rightarrow |e\rangle$ transition, e.g.,
following the heralded scheme \cite{Scully2006,Scully2009} (the case of more excitations is discussed below).
Applying optical fields propagating perpendicular to the emitter layer, see Fig \ref{fig1}(a), the collectively shared excitation has no phase variation across the ensemble.
\item A train of $(2 n_p +1)$ $\pi$-pulses resonant with
the $|e\rangle$-$|s\rangle$ transition bounces the state amplitude of the emitters back and forth between $|e\rangle$ and $|s\rangle$, while the in-plane wave-vectors
$\mathbf{k}_1$ or $\mathbf{k}_2$, see Fig. \ref{fig1}(b), cause accumulation of a wave vector that we design to
satisfy the equality $-(n_p+1)\mathbf{k}_1+n_p\mathbf{k}_2=\mathbf{k}_{sp}$.
\end{enumerate}
The combination of these processes produces the desired timed-Dicke state $|\psi_{\mathbf{k}_{sp}}\rangle$ \cite{Wang:2014aa}.
For schemes based on emissions from $|\psi_{\mathbf{k}_{sp}}\rangle$, two apparent contradictory requirements 
must be addressed: To
make the superradiant emission dominate the incoherent emissions, the ensemble should be optically thick for the
emitted mode \cite{Berman2011,Roehlsberger2010,Roof2016}, while the presumed uniform optical excitation
requires the ensemble to be optically thin during the state preparation \cite{Scully2007}.
We can indeed satisfy both conditions simultaneously:
The optical processes are either driven orthogonally
to the thin emitter ensemble or they act on only a single emitter population
(of states $\ket{e}$ and $\ket{s}$), thus the system is optically thin.
The emission modes here are SPP modes propagating parallel to the emitter layer, and
the SPP-emitter interaction is collectively enhanced by the large number of atoms in the final
internal state $\ket{g}$. Thus, the system may be optically thick upon emission.

The state $|e\rangle$ is coupled to $|g\rangle$ by a two-photon Raman process, and the collectively
shared excitation in $|e\rangle$ may hence be stable against spontaneous decay to the atomic ground state $|G\rangle=|g_1,g_2,\cdots g_N\rangle$.
During \emph{Step 2}, until we have completed the pulse train,
the intermediate timed-Dicke states
$|\psi_{-(n+1)\mathbf{k}_{1}+n\mathbf{k}_{2}}\rangle$ ($n<n_p$) have energy $\hbar\omega_{sg}$ but wavenumber smaller than $k_{sp}$ while larger than
the free-space resonant wavenumber $\omega_{sg}/c$.
These intermediate states may be protected from
decaying and emitting to SPP or free-space fields due to the wave number mismatch. Thus our
scheme works if the intermediate state lifetime supplies
enough time window for the $\pi$-pulses.

The length of the pulse train depends on the ratio between the wavelength of optical photons ($\lambda_{es}$) and
the SPP wavelength ($\lambda_{sp}$), $2n_p+1 \simeq \lambda_{es}/\lambda_{sp}$.
For the values of $\lambda_{sp}$,
graphene SPP may serve as an example.
Graphene SPPs are
distinguished by their tight confinement and long lifetime, and
by their high tunability via electrostatic gating \cite{Koppens2011,Grigorenko2012}.
For SPPs with frequency $\hbar\omega<2E_f$ \cite{GarciadeAbajo2014}
where $E_f\leq 1 \mathrm{eV}$ is the Fermi energy,
the graphene surface conductivity is approximated by the Drude conductivity: $\sigma_g(\omega)\approx i\frac{e^2}{\pi\hbar^2}E_f/(\omega+i\tau_D^{-1})$.
The value of $\tau_D$ currently available in experiments is $0.5\,\mathrm{ps}$ \cite{Woessner2014},
while it may intrinsically reach values of $10^2\,\mathrm{ps}$ \cite{Principi2013}.
Supposing for simplicity a vacuum below and above the graphene monolayer,
the dispersion relation of the p-mode graphene SPP is
$\omega_{sp}=\sqrt{2\alpha cE_f k_{sp}/\hbar}$,
where $\alpha\approx 1/137$ is the fine-structure constant.
Supposing $E_f=0.5\,$eV, then for $\hbar\omega_{sp}\in[0.01\,\mathrm{eV},
1\,\mathrm{eV}]$ \cite{Principi2013} $\lambda_{sp}$ ranges from 90 $\mathrm{\mu m}$ to 18 nm.

For optical pulses $\lambda_{es}\in [380\,\mathrm{nm},
750\,\mathrm{nm}]$, the number of pulses $2n_p+1\simeq\lambda_{es}/\lambda_{sp}<50$ and
even a single pulse is sufficient for low-energy SPPs with $\lambda_{sp}>\lambda_{es}$.
We can drive the optical $\pi$-pulses on the time scale of
nanoseconds using pulse powers that are far from damaging graphene \cite{Kiisk2013} and other surface polariton systems.

The validity of this scheme replies on details of the collective emitter-SPP coupling to be analyzed
in the following. We will first focus on the emission from $|\psi_{\mathbf{k}_{sp}}\rangle$.
Then we will study the decay of the intermediate states, which should be suppressed in order 
to successfully prepare $|\psi_{\mathbf{k}_{sp}}\rangle$. Our analysis will clarify the proper regime for the
experimental parameters.

\emph{Emitter-SPP Coupling}-To study the
emission from the prepared state $|\psi_{\mathbf{k}_{sp}}\rangle$, we now turn to the
coupling to the dispersive and dissipative electric field quantized as \cite{Dung1998,Philbin2010,Gruner1996}
\begin{equation}\label{e-field}
\begin{aligned}
  \mathbf{E}(\mathbf{r}_a) =i & \mu_0 \sqrt{\frac{\hbar\epsilon_0}{\pi}}
  \int_{0}^{\infty}d\tilde{\omega}\int d^3\mathbf{r}'
   \tilde{\omega}^2\sqrt{\Im\epsilon(\mathbf{r}',\tilde{\omega})}\\
  & \times  \mathbf{G}(\mathbf{r}_a,\mathbf{r}',\tilde{\omega})
  \cdot\mathbf{f}(\mathbf{r}',\tilde{\omega})
  +h.c.,
\end{aligned}
\end{equation}
where $\mu_0$ and $\epsilon_0$ are the
vacuum susceptibility and permittivity; $\Im\epsilon(\mathbf{r}',\tilde{\omega})$ is the imaginary
part of the relative permittivity; $\mathbf{G}(\mathbf{r},\mathbf{r'},\tilde{\omega})$ is
the dyadic Green's tensor determined by Maxwell's equations, and
the field $\mathbf{f}(\mathbf{r}',\tilde{\omega})$ with three
Cartesian operator components $f_j$ obeys the bosonic commutator relations
$[f_j, f_k]=0$, $[f_j^\dagger, f_k^\dagger]=0$ and
$[f_{j}(\mathbf{r}_1,\tilde{\omega}_1), f^\dagger_{k}(\mathbf{r}_2,\tilde{\omega}_2)]=
\delta_{jk}\delta(\mathbf{r}_1-\mathbf{r}_2)\delta(\tilde{\omega}_1-\tilde{\omega}_2)$.

The Hamiltonian is written as $H=\sum_{a=1}^{N}[\frac{1}{2}\hbar\omega_{sg}\sigma_a^z-
\sigma_a^x\mathbf{d}_a\cdot\mathbf{E}(\mathbf{r}_a)]+\int d^3\mathbf{r'}\int_{0}^{\infty}
d\tilde{\omega}\,\hbar\tilde{\omega}
\mathbf{f}^\dagger(\mathbf{r'},\tilde{\omega})\mathbf{f}(\mathbf{r'},\tilde{\omega})$
where $\mathbf{d}_a$ is the dipole of
the $|g_a\rangle$-$|s_a\rangle$ transition,
$\sigma^z_a=|s_a\rangle\langle s_a|-|g_a\rangle\langle g_a|$ and
$\sigma^x_a=|s_a\rangle\langle g_a|+|g_a\rangle\langle s_a|$. Here and throughout, $\hbar=1$.
We shall use the rotating-wave approximation and study the evolution based on the ansatz
written with time-dependent amplitudes $\alpha_a$ and $\beta_a(\tilde{\omega},\mathbf{r}')$:
\begin{equation}\label{ansatz}
|\Psi \rangle= \sum_{a=1}^{N}\alpha_a |s_a, {\varnothing} \rangle\bigotimes_{b\neq a}|g_{b}\rangle+
\int_{j,\tilde{\omega},\mathbf{r}'}\beta_j(\tilde{\omega},\mathbf{r}')
|G,1_{j,\tilde{\omega},\mathbf{r}'}\rangle
\end{equation}
where $|{\varnothing}\rangle$ is the field vacuum state,
$|1_{j,\tilde{\omega},\mathbf{r}'}\rangle=f_j^\dagger(\mathbf{r}',\tilde{\omega})|{\varnothing}\rangle$
and $\int_{j,\tilde{\omega},\mathbf{r}'}$ is the
short hand for $\sum_j\int d\tilde{\omega}\int d^3\mathbf{r}'$.
The strength of the emitter-emitter coupling
mediated by all environmental modes
\begin{equation}\label{gij}
  g_{ab}(\tilde{\omega})= \frac{\mu_0}{\pi}\tilde{\omega}^2\mathbf{d}_a\cdot
  \mathbf{G}(\mathbf{r}_a,\mathbf{r}_b,\tilde{\omega})\cdot\mathbf{d}_b
\end{equation}
has the symmetry $g_{ab}(\tilde{\omega})=g_{ba}(\tilde{\omega})$. Due to the
in-plane translation symmetry (we assume that the
dipoles of the emitters are identical)\cite{Novotny2012},
$g_{ab}(\tilde{\omega})$ can be expanded in the wave number representation
\begin{equation}\label{gk}
  g_{ab}(\tilde{\omega})=\int \frac{\mathbf{d^2k_{\shortparallel}}}{(2\pi)^2}\,
  g_{z_a, z_b}(\tilde{\omega},\mathbf{k_{\shortparallel}})
  e^{i\mathbf{k_{\shortparallel}}\cdot(\mathbf{r}_a-\mathbf{r}_b)},
\end{equation}
where the subindex ``$z_a, z_b$'' indicates the
emitter heights above the interface.
For a thin emitter layer, we approximate all the emitter $z$-coordinates by a single value
$z_{at}$, and we thus express $g_{z_a,z_b}(\tilde{\omega},\mathbf{k_{\shortparallel}})$ as $g_{z_{at}}(\tilde{\omega},\mathbf{k_{\shortparallel}})$.

Similarly, the excitation amplitudes of the individual emitters defined in Eq. (\ref{ansatz})
can also be transformed into wave number representation, i.e.,
$\alpha_{\mathbf{k}_{\shortparallel}}(t)=
\langle\psi_{\mathbf{k}_{\shortparallel}},{\varnothing}|\Psi(t)\rangle$,
which follows the equation
\begin{equation}\label{eqalpha}
\begin{aligned}
  -\partial_t\alpha_{\mathbf{k}_{\shortparallel}} & (t)=N \int\frac{d^2\mathbf{q}_{\shortparallel}}{(2\pi)^2}
  \int_{\tilde{\omega}}\Im g_{z_{at}}(\tilde{\omega},\mathbf{q}_{\shortparallel})
  \zeta(\mathbf{k}_{\shortparallel},\mathbf{q}_{\shortparallel}) \\
  &\quad\times\int_{0}^{t}d\tau \alpha_{\mathbf{q}_{\shortparallel}}(\tau)e^{-i(\tilde{\omega}-\omega_{sg})(t-\tau)}
  \end{aligned}
\end{equation}
where $\Im$ denotes the imaginary part and
\begin{equation}\label{zeta}
\zeta(\mathbf{q}_{\shortparallel}, \mathbf{k}_{\shortparallel})=
  \langle\psi_{\mathbf{q}_{\shortparallel}}|\psi_{\mathbf{k}_{\shortparallel}}\rangle
\end{equation}
is a geometry factor which quantifies the sharpness of the phase matching condition given the
spatial distribution of the emitters.
If $N\gg 1$ and the emitters are for example distributed independently according to a Gaussian distribution with width $L$,
$\zeta(\mathbf{k}_\shortparallel,\mathbf{q}_\shortparallel)=e^{-L^2(\mathbf{k}_\shortparallel-
\mathbf{q}_\shortparallel)^2/2}$.

The factor of $N$ in Eq. (\ref{eqalpha}) demonstrates the effect of collective enhancement. The collective
Lamb shift of state $|\psi_{\mathbf{k}_{sp}}\rangle$ should be considered unless it is smaller than the line width of the SPP mode.
For completeness, we provide the expressions for the collective Lamb shift in the Supporting Information.

\begin{figure}[b]
    \includegraphics[width=\textwidth]{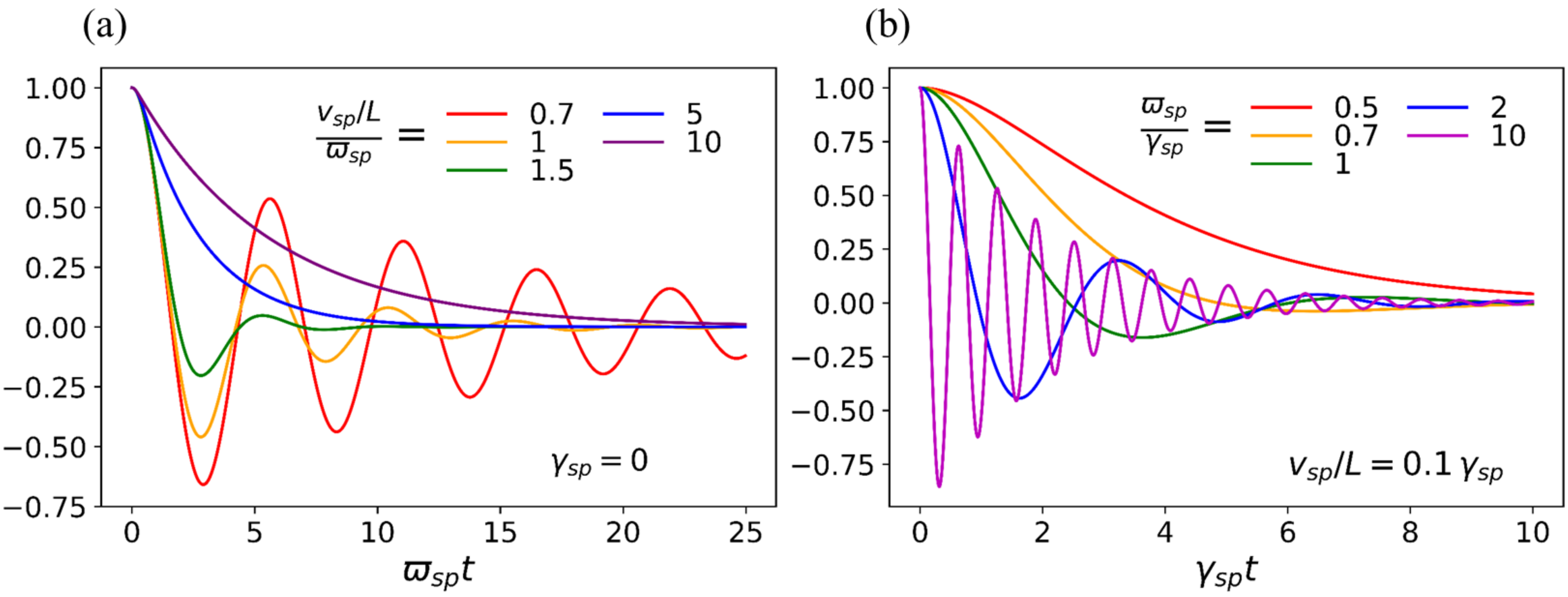}
    \caption{Evolution of $\alpha_{\mathbf{k}_{sp}}(t)$ determined by Eq. (\ref{eqksp2}).
    (a) $\varpi_{sp}$ and $\lambda_{sp}$ are fixed while $v_{sp}/L$ is varied.
    (b) $v_{sp}/L$ and $\gamma_{sp}$ are fixed while
    $\varpi_{sp}$ is varied.}
    \label{fig2}
\end{figure}

\emph{Evolution of $\ket{\psi_{\mathbf{k}_{sp}}}$}-To analyze the evolution described by Eq. (\ref{eqalpha}), we
shall start from the initial state $|\psi_{\mathbf{k}_{sp}},\varnothing\rangle$ and focus on the state amplitude $\alpha_{\mathbf{k}_{sp}}(t)$.
$|\psi_{\mathbf{k}_{sp}}\rangle$ resonantly matches the
SPP with frequency $\omega_{sg}$ while phase-matched photon modes are
off-resonant. Thus we may consider only the coupling to a range of SPPs. We use
$\omega_{\mathbf{k}_{\shortparallel}}$ and $\gamma_{\mathbf{k}_{\shortparallel}}$
to denote the frequency and damping rate of the SPP with in-plane momentum $\mathbf{k}_{\shortparallel}$.
They are determined by the position of the
pole of $g_{z,z'}(\tilde{\omega},\mathbf{k}_{\shortparallel})$ in the
complex $\tilde{\omega}$ plane \cite{Novotny2012}. Keeping only the contribution from the poles
leads to a Lorentzian type expression
\begin{equation}\label{lorentz}
  \Im g_{z, z'}(\tilde{\omega},\mathbf{k}_{\shortparallel})\approx
  \frac{A_{z, z'}(\mathbf{k}_{\shortparallel})\gamma_{\mathbf{k}_{\shortparallel}}}
  {(\tilde{\omega}-\omega_{\mathbf{k}_{\shortparallel}})^2+\gamma_{\mathbf{k}_{\shortparallel}}^2},
\end{equation}
where $A_{z,z'}(\mathbf{k}_{\shortparallel})$
is fixed by the residue of $g_{z,z'}(\tilde{\omega},\mathbf{k}_{\shortparallel})$ at the pole
$\tilde{\omega}=\omega_{\mathbf{k}_{\shortparallel}}-i\gamma_{\mathbf{k}_{\shortparallel}}$.

When $\zeta(\mathbf{k}_{\shortparallel},\mathbf{k}_{sp})$
peaks sharply at $\mathbf{k}_{\shortparallel}=\mathbf{k}_{sp}$, the distribution of the emitter excitation
is centered at $|\psi_{\mathbf{k}_{sp}}\rangle$ so that
$\alpha_{\mathbf{k}_{\shortparallel}}(t)\approx\langle\psi_{\mathbf{k}_{\shortparallel}}
|\psi_{\mathbf{k}_{sp}}\rangle\langle\psi_{\mathbf{k}_{sp}},{\varnothing}|\Psi\rangle
=\alpha_{\mathbf{k}_{sp}}(t)\zeta(\mathbf{k}_{\shortparallel},\mathbf{k}_{sp})$.
This approximation makes it possible to obtain a closed equation of evolution for
$\alpha_{\mathbf{k}_{sp}}$, which, with the Gaussian distribution of emitters and the corresponding geometry factor, is written as
\begin{equation}\label{eqksp2}
  -\partial_t\alpha_{\mathbf{k}_{sp}}=\varpi_{sp}^2
  \int_{0}^{t}d\tau\, \alpha_{\mathbf{k}_{sp}}(\tau)
  e^{-\frac{\mathbf{v}_{sp}^2}{4L^2}(t-\tau)^2-\gamma_{sp}(t-\tau)},
\end{equation}
where $\varpi_{sp}^2=\frac{N}{4L^2}A_{z_{at}}(\mathbf{k}_{sp})$, and
$\mathbf{v}_{sp}=\nabla_{\mathbf{k}_{\shortparallel}}\omega_{\mathbf{k}_{\shortparallel}}
|_{\mathbf{k}_{\shortparallel}=\mathbf{k}_{sp}}$ is the SPP group velocity.
Unlike the case of free-photon superradiance \cite{Svidzinsky2008},
here the finite SPP lifetime due to Ohmic damping must be considered. We assume the SPP decay rate as a
constant $\gamma_{\mathbf{k}_{\shortparallel}}=\gamma_{sp}$. For the Drude model of graphene mentioned above,
$\gamma_{sp}=0.5\,\tau_D^{-1}$. See the Supporting Information for the derivation of Eq. (\ref{eqksp2}).

The solution to Eq. (\ref{eqksp2}) behaves as damped oscillations or pure decay
depending on the interplay between three parameters, viz.,
$\varpi_{sp}$, $\gamma_{sp}$ and $v_{sp}/L$.
The damped oscillation regime appears when $\varpi_{sp}$, which
is roughly the frequency of the oscillation, dominates the other two parameters.
To understand this condition, note that the oscillation refers to the periodical absorption and reemission of
the single-SPP pulse. This process is possible only until the propagating pulse leaves
the ensemble at $t\simeq L/v_{sp}$ or has been absorbed by the
material due to Ohmic loss at $t\simeq\gamma_{sp}^{-1}$.
In Fig. \ref{fig2}(a) we show the solution of Eq. (\ref{eqksp2}) as function of time for different values of the finite duration $L/v_{sp}$ of the SPP pulse propagation in the emitter ensemble.
Figure \ref{fig2}(b) shows the similar results when the damping is mainly determined by the finite
SPP lifetime, $\gamma_{sp}^{-1}$. Both plots of Fig. \ref{fig2} confirm the role of
$\varpi_{sp}$ as oscillation frequency.
In the pure decay regime ($\varpi_{sp}<\gamma_{sp},v_{sp}/L$), we obtain the Markov
approximation by assuming
$\alpha_{\mathbf{k}_{sp}}(\tau)=\alpha_{\mathbf{k}_{sp}}(t)$ in
the right hand side of Eq. (\ref{eqksp2}). It yields the decay rate
\begin{equation}\label{decay}
  \Gamma_c=\sqrt{\pi}\frac{\varpi_{sp}^2}{v_{sp}/L}e^{(\frac{\gamma_{sp}}{v_{sp}/L})^2}
  \mathrm{erfc}(\frac{\gamma_{sp}}{v_{sp}/L}).
\end{equation}
This expression verifies our observations in Fig. \ref{fig2}, e.g., that larger $\varpi_{sp}$ and
$v_{sp}/L$ result in faster decay.
When $v_{sp}/L$ can be neglected, $\Gamma_c\approx \varpi_{sp}^2/\gamma_{sp}<\gamma_{sp}$.

Both the damped oscillation and pure decay regimes are achievable in experiments.
With realistic parameters $\omega_{sg}=E_f=0.5\,\mathrm{eV}$,
the SPP group velocity is roughly $v_{sp}=10^{-2}c$.
so that for $L\geq 1\,\mathrm{\mu m}$, $v_{sp}/L \leq 10^{12}\,\mathrm{Hz}=\gamma_{sp}$.
For emitter vacuum decay rate $10^2\,\mathrm{MHz}$ (governed by the transition dipole moment),
emitter-graphene distance $z_{at}=10\,\mathrm{nm}$
and emitter number density $n_{at}=N/L^2=(0.1/\mathrm{nm})^{2}$,
the damped oscillation regime is reached with $\varpi_{sp}\approx 10^{14}\,\mathrm{Hz}\gg \gamma_{sp}, v_{sp}/L$, see Methods.
The pure decay
regime can be realized by larger distance $z_{at}$, lower density $n_{at}$, or a smaller transition dipole moment.

\begin{figure}[t]
    \includegraphics[width=\textwidth]{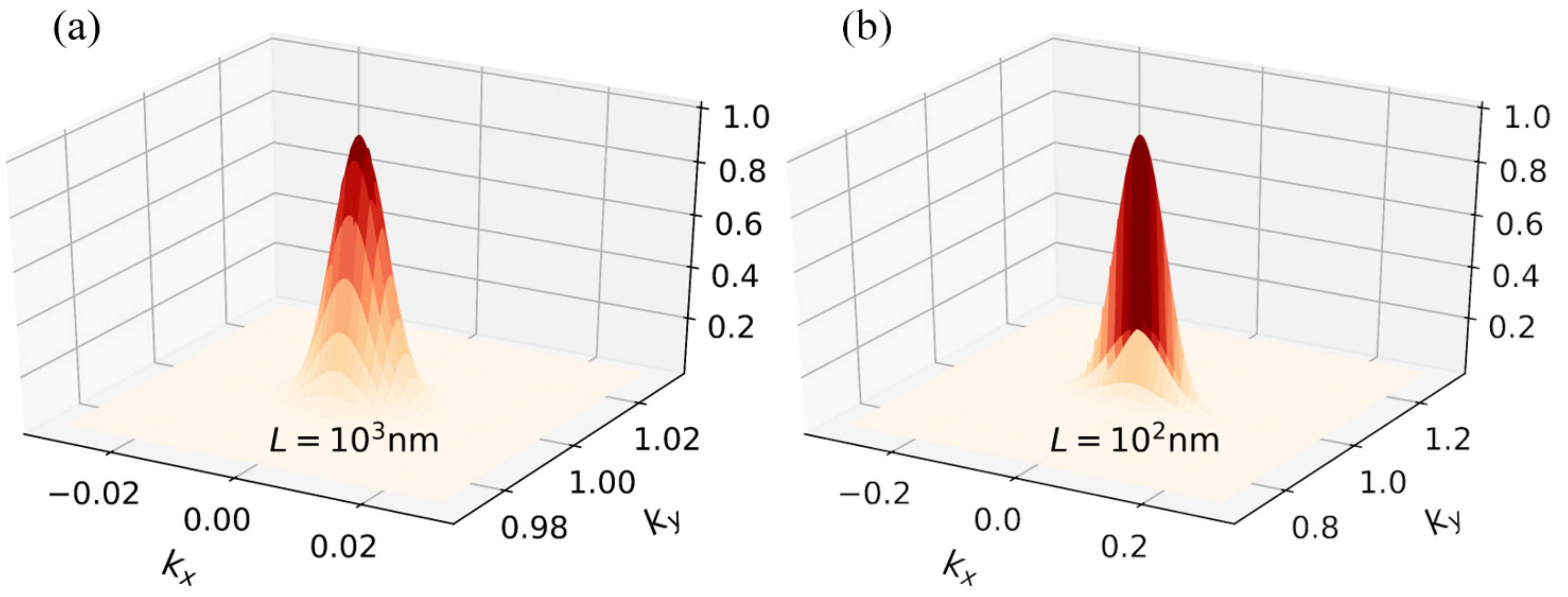}
    \caption{$P(\mathbf{k}_{\shortparallel}, \omega_{sp})/P(\mathbf{k}_{sp}, \omega_{sp})$
     for graphene SPPs when $\omega_{sp}=E_f=0.5\,\mathrm{eV}$, $\tau=1\,\mathrm{ps}$, emitter-graphene
     distance $z_{at}=10\,\mathrm{nm}$, ensemble size (a) $L=10^3\,\mathrm{nm}$; (b) $L=10^2\,\mathrm{nm}$.
     $\mathbf{k}_{sp}$ is set to along $\hat{y}$-direction. The unit of the wave number is
     $k_{sp}$ of which the SPP wavelength is $\lambda_{sp}=36.2\,\mathrm{nm}$.}
    \label{fig3}
\end{figure}

\emph{Directionality of the emitted SPP}-
The amplitudes $\beta_j(\tilde{\omega},\mathbf{r}')$ defined in Eq. (\ref{ansatz}) can be transformed
into wave number representation $\beta_j(\tilde{\omega},\mathbf{k}_{\shortparallel},z')$.
Although exact solutions are not accessible, we may assume a uniform decay ansatz
$\alpha_a(t)=\alpha_a(0)e^{-\gamma t}$, with which
the electromagnetic frequency-wave number excitation distribution
$P(\mathbf{k}_{\shortparallel},\tilde{\omega})\equiv\int dz'
\sum_j |\beta_j(\tilde{\omega},\mathbf{k}_{\shortparallel},z')|^2$, is
\begin{equation}\label{direction}
P(\mathbf{k}_{\shortparallel},\tilde{\omega})
=\frac{N \Im g_{z_{at}}(\tilde{\omega},\mathbf{k}_{\shortparallel})
|\zeta(\mathbf{k}_{sp},\mathbf{k}_{\shortparallel})|^2}{\gamma^2+(\tilde{\omega}-\omega_{sg})^2}.
\end{equation}
Both the pole structure of
$g_{z}(\tilde{\omega},\mathbf{k}_{\shortparallel})$ and the geometry factor
$\zeta(\mathbf{k}_{sp},\mathbf{k}_{\shortparallel})$ in this formula guarantee the emission to peak sharply at
$\mathbf{k}_{\shortparallel}=\mathbf{k}_{sp}$.
The ratio between $P(\mathbf{k}_{\shortparallel},\omega_{sp})$
and the peak value $P(\mathbf{k}_{sp},\omega_{sp})$ is depicted in Fig. \ref{fig3}.
The figure confirms that the larger ensemble size leads to stronger directionality.

\begin{figure}[b]
    \includegraphics[width=\textwidth]{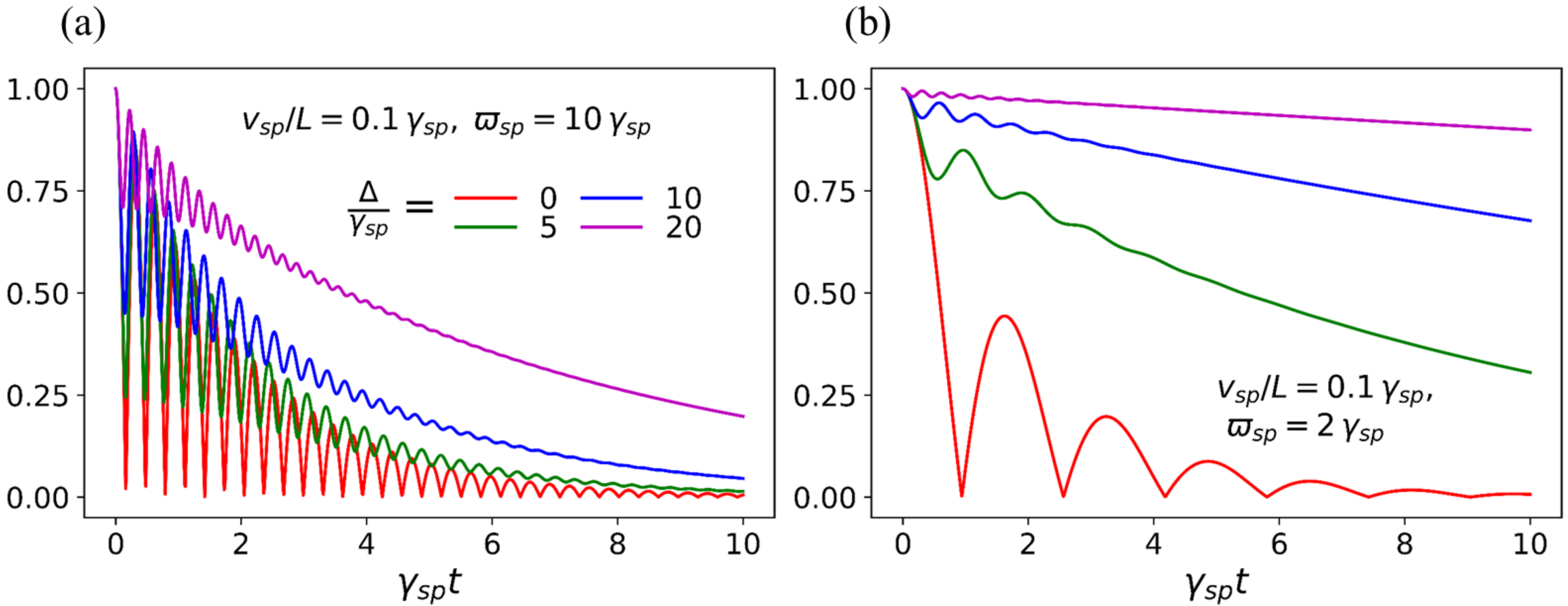}
    \caption{Evolution of $|\alpha_{\mathbf{q}_{n}}(t)|$ of the intermediate states with varying
    values of $\Delta$. $v_{sp}/L=0.1\,\gamma_{sp}$, (a)$\varpi_{sp}=10\,\gamma_{sp}$, (b)$\varpi_{sp}=2\,\gamma_{sp}$. }
    \label{fig4}
\end{figure}

\emph{Evolution of Intermediate States}-Now we turn to the intermediate states
$|\psi_{\mathbf{q}_{n}}\rangle$ that may be
populated for nanoseconds during \emph{Step 2} in the preparation of $|\psi_{\mathbf{k}_{sp}}\rangle$, where $\mathbf{q}_{n}= -(n+1)\mathbf{k}_{1}
+n\mathbf{k}_{2}$ ($0\leq n<n_p$). Since $\omega_{es}>\omega_{sg}$ and
$\omega_{sg}/c<q_{n}<k_{sp}$, the photon and SPP modes matching the wave vector $q_{n}\simeq (2n+1)\omega_{es}/c$ 
are not resonant with $\omega_{sg}$. We denote the two detunings as $\Delta_n=\omega_{sg}-\omega_{\mathbf{q}_{n}}$ and
$\Delta_n^{(0)}= cq_{n}-\omega_{sg}$, respectively.
For $\omega_{es}$ in the optical regime
and graphene SPP frequency $\omega_{sg}$ at most in the near-infrared,
$\Delta_n^{(0)}\gg\Delta_n$ ($\Delta_n^{(0)}$ ranges from optical to ultraviolet frequencies) thus we can disregard the coupling
to free-space photons. Then the equation of evolution for $\alpha_{\mathbf{q}_{n}}(t)$
resembles Eq. (\ref{eqksp2}) but with the replacement
$\gamma_{sp}\rightarrow\gamma_{sp}+i\Delta_n$ and $\varpi_{sp}, v_{sp}$
should be defined by the wave vector $\mathbf{q}_n$ (see the Supplemental Material for the
full formulae). In the damped oscillation regime, we illustrate the
solution to $|\alpha_{\mathbf{q}_{n}}|$ in Fig. \ref{fig4}. It shows that compared with the
phase-matched case ($\Delta_{n_p}=0$), the intermediate states
have longer lifetime, especially when $\varpi_{sp}$ is smaller.

In the pure decay regime, the Markov approximation yields the decay rate
of $|\psi_{\mathbf{q}_{n}}\rangle$,
\begin{equation}\label{decay-in}
  \Gamma_{\mathbf{q}_{n}}=\frac{\varpi_{sp}^2}{\gamma_{sp}^2+\Delta_n^2}\gamma_{sp},
\end{equation}
where we have omitted terms of order $v_{sp}/L$ that are dominated by $\gamma_{sp}$ in a large ensemble.
Since $\gamma_{sp}$ is the uncertainty of the SPP frequency, in practice we would require $\Delta_n\gg\gamma_{sp}$.
For near-infrared graphene SPP with $\hbar\omega_{sp}=E_f=0.5\,\mathrm{eV}$
and wavelength $\lambda_{sp}=36.2\,\mathrm{nm}$, supposing optical pulse wavelength
$\lambda_{es}=500\,\mathrm{nm}$, 15 pulses are sufficient to prepare $\ket{\psi_{\mathbf{k}_{sp}}}$ and
$\Delta_n\geq\Delta_{n_p-1} \approx 0.035\,\mathrm{eV}\approx 10\gamma_{sp}$ for all intermediate states
$\ket{\psi_{\mathbf{q}_n}}$ with $n<n_p$. 
Equation (\ref{decay-in}) implies that when $\varpi_{sp}< 0.1\gamma_{sp}$, $|\psi_{\mathbf{q}_{n_p-1}}\rangle$
has a lifetime longer than $10^4\gamma_{sp}^{-1}=10\,\mathrm{ns}$, allowing its population for $\leq$ 1 $\mathrm{ns}$
during the last two $\pi$-pulses required to prepare $\ket{\psi_{\mathbf{k}_{sp}}}$.

In the damped oscillation regime, however, the lifetime of the intermediate states may be
too short to facilitate the preparation of $|\psi_{\mathbf{k}_{sp}}\rangle$. In this case, we may employ another
metastable emitter level $|s'\rangle$ which disallows the direct $\ket{s'}$-$\ket{g}$ transition. The SPP wave number is then
accumulated with the $\ket{e}$-$\ket{s'}$ transition and a final pulse moves the
collective excitation from $\ket{s'}$ to $\ket{s}$ to obtain $\ket{\psi_{\mathbf{k}_{sp}}}$.

\emph{Conclusions and Discussions}-We have proposed to prepare an emitter ensemble into collective states
that match the wave vector of surface plasmon polaritons and hence
directionally emit SPPs via polariton superradiance. The directionality has high tunability, i.e., the direction is determined simply by the
wave vectors of the $\pi$-pulses used in the preparation of the timed-Dicke states. 
We studied the evolution of the collective emitter excitation and showed that the intermediate states
have lifetime long enough to implement the required pulses sequence.
With the Drude model parameters of graphene SPP in the near-infrared regime,
we predict excellent directionality of launching.
Our main general formalism also applies to other families of surface
polaritons \cite{Hillenbrand2002,Taubner2006,Dai2014,Low2016,Li2018,Basov2016,Lin2017,Woessner2014}.

In the end, we emphasize that the requirement of only a
single excitation in the timed-Dicke state can be released to the low-excitation regime.
In this regime, the spin operators can be approximated with bosonic ladder operators
$\sigma^-_a\rightarrow \hat{b}_a$. We can translate them into momentum representation by
$\hat{b}_{\mathbf{q}_{\shortparallel}}=\frac{1}{\sqrt{N}}\sum_{a=1}^{N}\hat{b}_a e^{-i\mathbf{q}_{\shortparallel}\cdot\mathbf{r}_a}$ with
$[\hat{b}_{\mathbf{q}_{\shortparallel}}, \hat{b}_{\mathbf{k}_{\shortparallel}}^\dagger]
=\zeta(\mathbf{q}_{\shortparallel},\mathbf{k}_{\shortparallel})\approx \delta_{\mathbf{q}_{\shortparallel},\mathbf{k}_{\shortparallel}}$,
The weak coherent state amplitudes then obey similar equations
as the single excitation amplitudes in Eq. (\ref{ansatz}) \cite{Porras2008}.
For a system with inhomogeneous broadening, for example, doped rare-earth ions in crystals, the
$\pi$-pulses should be implemented in a more sophisticated way \cite{Remizov:2015aa}. 
The influence of inhomogeneous broadening and dephasing on phase coherence has been studied in other contexts \cite{Moiseev2001} while the effect on 
the SPP emission may be minor because SPP modes already have broad bandwidth in the range of THz. Besides the applications of directional SPPs
\cite{Lopez-Tejeira2007,Lin2013,Pors2014,You2015,Krasnok2018,Bliokh2018,Song2017,Andolina2018},
our results may also facilitate interfaces between
photonic and plasmonic systems for quantum information processing \cite{Tame2013,Bozhevolnyi2018}.
Other phenomena related
to single-photon superradiance, such as superradiance amplification \cite{Svidzinsky2013} and
superradiance lattice \cite{Wang2015}, may also be investigated in surface polariton systems
based on our scheme for timed-Dicke states.

\section{Methods}
The surface conductivity of the graphene monolayer
given by the Drude model is
\begin{equation}\label{drude}
  \sigma(E_f,\tau; \tilde{\omega})= \frac{e^2E_f}{\pi\hbar^2}\frac{i}{\tilde{\omega}+i\tau^{-1}_D}.
\end{equation}
This expression is convenient for
our analysis
when the temperature is low and $\hbar\tilde{\omega}<2E_f$.
For the graphene layer, the Fresnel coefficient of reflection of the p-modes is
\begin{equation}
  r_p(\tilde{\omega},\mathbf{q}_{\shortparallel})=\frac{\sigma(\tilde{\omega})q_{z}}
  {2\tilde{\omega}\epsilon_0+\sigma(\tilde{\omega})q_{z}},
\end{equation}
where $q_{z}=\sqrt{\tilde{\omega}^2/c^2-\mathbf{q}_{\shortparallel}^2}$ and for graphene SPPs
$q_z\approx iq_{\shortparallel}$ \cite{Koppens2011}. In the above expression, we have assumed
that the dielectrics above and below the graphene monolayer are vacuum.

When the emitters are polarized perpendicular to the graphene layer,
only one element of the scattering part of the dyadic Green's tensor is relevant,
which yields the coupling strength
\begin{equation}
  g_{z,z'}(\mathbf{q}_{\shortparallel},\tilde{\omega})=
  \frac{i}{2\pi\epsilon_0 q_{z}}|\mathbf{d}|^2 q_{\shortparallel}^2 r_p(\tilde{\omega},\mathbf{q}_{\shortparallel}) e^{-q_{\shortparallel} (z+z')},
\end{equation}
The poles that define the SPP are given by the equation
\begin{equation}
  \tilde{\omega}(\tilde{\omega}+i\tau_D^{-1})=2\alpha cE_f q_{\shortparallel}/\hbar,
\end{equation}
where $\alpha\approx 1/137$ is the fine structure constant.
Solutions of the above equation imply that
$\omega_{\mathbf{q}_{\shortparallel}}=\sqrt{2\alpha c E_f q_{\shortparallel}/\hbar}$ and $\gamma_{\mathbf{q}_{\shortparallel}}=0.5\,\tau_D^{-1}$
when $\omega_{\mathbf{q}_{\shortparallel}}\gg\gamma_{\mathbf{q}_{\shortparallel}}$. Indeed, if the condition
$\omega_{\mathbf{q}_{\shortparallel}}\gg\gamma_{\mathbf{q}_{\shortparallel}}$ is not satisfied, the Drude model conductivity
should be replaced with more advanced expressions to yield well-defined SPPs.

The residue of $r_p$ at the pole $\tilde{\omega}=\omega_{\mathbf{q}_{\shortparallel}}-i\gamma_{\mathbf{q}_{\shortparallel}}$ is
$0.5\,\omega_{\mathbf{q}_{\shortparallel}}$, and $A_{z,z'}(\mathbf{q}_{\shortparallel})$ defined in Eq. (8) of the main text is given as
\begin{equation}
  A_{z,z'}(\mathbf{q}_{\shortparallel})=\frac{3\hbar\gamma_0}{4\omega_{sg}^3}c^3
  \omega_{\mathbf{q}_{\shortparallel}}q_{\shortparallel}e^{-q_{\shortparallel}(z+z')},
\end{equation}
where we have used the vacuum spontaneous emission rate $\gamma_0$ to express the
transition dipole.

For $E_f= 0.5\,\mathrm{eV}$,
the SPP wave number is $q_{sp}=0.174\,\mathrm{nm}^{-1}$ when $\hbar\omega_{sp}=0.5\,\mathrm{eV}$. Suppose that the distance between
the emitter layer and the graphene layer is $z_{at}=10\,\mathrm{nm}$. Then
we obtain $A_{z_{at}}=1.87\times 10^{20}\gamma_0\,\mathrm{(nm)^2/s}$. For
larger distance, e.g., $z_{at}=20\,\mathrm{nm}$, $A_{z_{at}}=5.73\times 10^{18}\gamma_0\,\mathrm{(nm)^2/s}$.

\emph{Supporting Information}
The supporting Information contains the derivation of Eq. (\ref{eqksp2}),
details of the analysis of intermediate timed-Dicke states dissipation, and the collective emission rate and the Lamb shift.

\section{Acknowledgement}
We sincerely thank Klaas-Jan Tielrooij for useful discussions and suggestions.
This work was supported by
European Union's Horizon 2020 research and innovation
program (No. 712721, NanOQTech) and the Villum Foundation.

\section{Supplemental Material}
\setcounter{equation}{0}
\renewcommand{\theequation}{S.\arabic{equation}}

In the Supplemental Material, we shall
present the derivation of Eq. (9) of the main text,
details of the analysis of intermediate timed-Dicke states dissipation, and the collective emission rate and the Lamb shift.

\subsection{A. Derivation of Eq. (9) of the Main Text}
The equations of evolution for the amplitudes introduced in Eq. (3) of the main text are
\begin{equation}\label{al1}
  -i\partial_t\alpha_a =\int_{j,\mathbf{r}',\tilde{\omega}}g_j(\mathbf{r}_a,\mathbf{r}',\tilde{\omega})
  \beta_{j}(\mathbf{r}',\tilde{\omega})e^{-i(\tilde{\omega}-\omega_{sg})t},
\end{equation}
and
\begin{equation}\label{beta1}
-i\partial_t\beta_j(\mathbf{r}',\tilde{\omega})=
\sum_{a=1}^{N}\alpha_a g_j^*(\mathbf{r}_a,\mathbf{r}',\tilde{\omega})e^{i(\tilde{\omega}-\omega_{sg})t}.
\end{equation}
Substituting Eq. (\ref{beta1}) into Eq. (\ref{al1}) yields
\begin{equation*}
  -\partial_t\alpha_a(t)=\sum_b\int_{0}^{\infty}d\tilde{\omega} \Im g_{ab}(\tilde{\omega})
  \int_{0}^{t}d\tau \alpha_b(\tau)e^{-i(\tilde{\omega}-\omega_{sg})(t-\tau)}.
\end{equation*}
Then we go to the wave number representation and assuming the Gaussian distribution of the
emitters as presented in the main text:
\begin{equation}
\begin{aligned}
  -\partial_t\alpha_{\mathbf{k}_{sp}}(t)=N & \int\frac{d^2\mathbf{q}_{\shortparallel}}{(2\pi)^2}\int d\tilde{\omega}\Im
  g_{z_{at}}(\tilde{\omega},\mathbf{q}_{\shortparallel})
  e^{-\frac{L^2}{2}(\mathbf{q}_{\shortparallel}-\mathbf{k}_{sp})^2} \\
  & \times\int_{0}^{t}d\tau \alpha_{\mathbf{q}_{\shortparallel}}(\tau)e^{-i(\tilde{\omega}-\omega_{sg})(t-\tau)}
  \end{aligned}
\end{equation}
We substitute the approximation of $g_{z_{at}}(\tilde{\omega},\mathbf{q}_{\shortparallel})$
introduced in Eq. (8) of the main text so that the integral over $\tilde{\omega}$ becomes
\begin{equation}\label{app-w}
\begin{aligned}
&   \int_0^\infty d\tilde{\omega}\;\Im
  g_{z_{at}}(\tilde{\omega},\mathbf{q}_{\shortparallel}) e^{-i(\tilde{\omega}-\omega_{sg})(t-\tau)} \\
 \approx & \pi A_{z_{at}}(\mathbf{q}_{\shortparallel})
e^{-i(\omega_{\mathbf{q}_{\shortparallel}}-i\gamma_{\mathbf{q}_{\shortparallel}}-\omega_{sg})(t-\tau)}.
\end{aligned}
\end{equation}
Then we substitute $\alpha_{\mathbf{q}_{\shortparallel}}\approx\alpha_{\mathbf{k}_{sp}}
\zeta(\mathbf{q}_{\shortparallel},\mathbf{k}_{sp})$ and
perform the integral over the in-plane momentum $\mathbf{q}_{\shortparallel}$.
We expand the expression as function of the surface plasmon
frequency around this peak and get the approximation that
\begin{equation*}
\begin{aligned}
& \int\frac{d^2\mathbf{q}_{\shortparallel}}{(2\pi)^2}\pi A_{z_{at}}(\mathbf{q}_{\shortparallel})
e^{-L^2(\mathbf{q}_{\shortparallel}-\mathbf{k}_{sp})^2
-i(\omega_{\mathbf{q}_{\shortparallel}}-i\gamma_{\mathbf{q}_{\shortparallel}}-\omega_{sg})(t-\tau)}\\
\approx &
\pi A_{z_{at}}(\mathbf{k}_{sp})
\int\frac{d^2\mathbf{p}_{\shortparallel}}{(2\pi)^2}e^{- L^2 p_{\shortparallel}^2
-i[\mathbf{p}_{\shortparallel}\cdot\mathbf{v}_{sp}-i\gamma_{\mathbf{k}_{sp}}](t-\tau)},
\end{aligned}
\end{equation*}
where we have made the substitution $\mathbf{q}_{\shortparallel}
\rightarrow\mathbf{p}_{\shortparallel}=\mathbf{q}_{\shortparallel}-\mathbf{k}_{sp}$,
and the surface plasmon frequency is expanded at $\mathbf{q}_{\shortparallel}=\mathbf{k}_{sp}$.
The integral over $\mathbf{p}_{\shortparallel}$ can be written as
\begin{equation}\label{t1}
\begin{aligned}
  & \int\frac{p_{\shortparallel} dp_{\shortparallel} d\theta}{(2\pi)^2}\; e^{-L^2 p_{\shortparallel}^2-
(iv_{sp}p_{\shortparallel}\cos\theta+\gamma_{sp})(t-\tau)} \\
= & \frac{1}{4\pi L^2}e^{-\frac{v_{sp}^2}{4L^2}(t-\tau)^2-\gamma_{sp}(t-\tau)},
\end{aligned}
\end{equation}
where we have assumed a constant SPP loss rate $\gamma_{sp}$.
Then we get Eq. (9) of the main text.

\subsection{B. Dissipation of the Intermediate Timed-Dicke States}
For the initial emitter state $|\psi_{\mathbf{k}_0}\rangle$ with $\mathbf{k}_0\neq \mathbf{k}_{sp}$,
the SPP channel may or may not dominate the emission into photon free-space photon modes.
The coupling strength to the free-space photon modes is
\begin{equation*}
  g^0(\tilde{\omega}, \mathbf{q}_{\shortparallel})=\frac{\mathbf{d}^2}{\pi\epsilon_0}\frac{i}{2q_z}\mathbf{q}_{\shortparallel}^2
\end{equation*}
where $q_{z}=\sqrt{\tilde{\omega}^2/c^2-\mathbf{q}_{\shortparallel}^2}$,
and we require that $\Im q_{z} \geq 0$.
The subsequent calculation follows the outline of the previous section.
\begin{equation}
\begin{aligned}
& \int_0^\infty d\tilde{\omega}\;\Im
  g^0(\tilde{\omega},\mathbf{q}_{\shortparallel}) e^{-i(\tilde{\omega}-\omega_{sg})(t-\tau)}  \\
= & -i\frac{c}{4\epsilon_0}\mathbf{d}^2 \mathbf{q}_{\shortparallel}^2 H^{(2)}_0[cq_\shortparallel(t-\tau)]
e^{i\omega_{sg}(t-\tau)},
\end{aligned}
\end{equation}
where $H^{(2)}_0$ is the zero-order Hankel function of the second kind.
To implement the integral over $\mathbf{q}_{\shortparallel}$, we write
$H^{(2)}_0(x)=\tilde{H}^{(2)}_0(x)e^{-ix}$
making use of the asymptotic behavior of Hankel functions.
We fix the slowly varying part $\tilde{H}^{(2)}_0(x)$ by its value at $\mathbf{q}_{\shortparallel}=\mathbf{k}_0$
and integrate only the fast oscillating phase factor $e^{-ix}$. This yields
\begin{equation}
\begin{aligned}
  & \int \frac{d^2\mathbf{q}_{\shortparallel}}{(2\pi)^2}e^{-L^2(\mathbf{q}_{\shortparallel}-\mathbf{k}_0)^2}
  \mathbf{q}_{\shortparallel}^2 H^{(2)}_0[\omega^0_{\mathbf{q}_\shortparallel}(t-\tau)]
e^{i\omega_{sg}(t-\tau)}\\
\approx  & \frac{\mathbf{k}_0^2}{4\pi L^2} \tilde{H}^{(2)}_0[ck_0(t-\tau)]
e^{-i\Delta_0(t-\tau)-\frac{c^2}{4L^2}(t-\tau)^2}.
\end{aligned}
\end{equation}
where $\Delta_0=ck_0-\omega_{sg}$.
Meanwhile, the contribution from SPPs, Eq. (\ref{t1}), will acquire an additional off-resonant factor,
\begin{equation}
  Eq. (\ref{t1})\rightarrow
  \frac{1}{4\pi L^2}e^{-\frac{v_{sp}^2}{4L^2}(t-\tau)^2-(\gamma_{sp}-i\Delta)(t-\tau)}.
  \end{equation}
where $\Delta=\omega_{sg}-\omega_{\mathbf{k}_0}$ is the detuning between the SPP with momentum
$\mathbf{k}_0$ and the emitter excitation.
The final equation for the amplitude $\alpha_{\mathbf{k}_0}$ is
\begin{equation}
\begin{aligned}
  -i\partial_t\alpha_{\mathbf{k}_0}(t) &= -i\varpi_0^2\int_{0}^{t}d\tau\,\alpha_{\mathbf{k}_0}(\tau)
  \tilde{H}^{(2)}_0[ck_0(t-\tau)]\\
 &\qquad\qquad  \times e^{-i\Delta_0(t-\tau)-\frac{c^2}{4L^2}(t-\tau)^2} \\
 + & \varpi_{sp}^2\int_{0}^{t}d\tau\,\alpha_{\mathbf{k}_0}(\tau)
  e^{-\frac{v_{sp}^2}{4L^2}(t-\tau)^2-(\gamma_{sp}-i\Delta)(t-\tau)},
\end{aligned}
\end{equation}
where $\varpi^2_0=\frac{N c \mathbf{d}^2\mathbf{k}_0^2}{(4\pi)^2\epsilon_0 L^2}$.

\subsection{C. Collective Emission Rate and Lamb Shift}
Our ansatz for the quantum state goes beyond the rotating-wave approximation
and is more general than Eq. (3) of the main text. With additional terms in the three-excitations manifold, the ansatz is written as
\begin{equation}
\begin{aligned}
|\Psi & \rangle= \sum_{a=1}^{N}\alpha_a |s_a, {\varnothing} \rangle\bigotimes_{b\neq a}|g_{b}\rangle+
\int_{j,\tilde{\omega},\mathbf{r}'}\beta_j(\tilde{\omega},\mathbf{r}')
|G,1_{j,\tilde{\omega},\mathbf{r}'}\rangle\\
& +\sum_{(a,b)}\int_{j,\tilde{\omega},\mathbf{r}'}
\xi_{ab,j}(\tilde{\omega},\mathbf{r}')|s_a, s_b, 1_{j,\tilde{\omega},\mathbf{r}'}\rangle\bigotimes_{c\neq a,b}|g_{c}\rangle,
\end{aligned}
\end{equation}
where $\sum_{(a,b)}$ means summation over pairs of $a\neq b$.
Equations for the time-dependent amplitudes of the above ansatz are given as
\begin{equation}
\begin{aligned}
  -i\partial_t\alpha_a & = \int_{j,\mathbf{r}',\tilde{\omega}}g_j(\mathbf{r}_a,\mathbf{r}',\tilde{\omega})
  \beta_{j}(\mathbf{r}',\tilde{\omega})e^{-i(\tilde{\omega}-\omega_{a,sg})t}+ \\
  & \sum_{b\neq a} \int_{j,\mathbf{r}',\tilde{\omega}}\xi_{ab,j}(\mathbf{r}',\tilde{\omega})
  g_j(\mathbf{r}_b,\mathbf{r}',\tilde{\omega})e^{-i(\omega_{b,sg}+\tilde{\omega})t},
\end{aligned}
\end{equation}
\begin{equation}\label{eqbeta}
-i\partial_t\beta_j(\mathbf{r}',\tilde{\omega})=
\sum_{a=1}^{N}\alpha_a g_j^*(\mathbf{r}_a,\mathbf{r}',\tilde{\omega})e^{i(\tilde{\omega}-\omega_{a,sg})t},
\end{equation}
and
\begin{equation}\label{eqxi}
\begin{aligned}
-i\partial_t\xi_{ab,j}(\mathbf{r}',\tilde{\omega})=&\alpha_a
g^*_j(\mathbf{r}_b,\mathbf{r}',\tilde{\omega})e^{i(\tilde{\omega}+\omega_{b,sg})t} \\
& +\alpha_b g^*_j(\mathbf{r}_a,\mathbf{r}',\tilde{\omega})e^{i(\tilde{\omega}+\omega_{a,sg})t}.
\end{aligned}
\end{equation}
In the above equations,
\begin{equation*}
  g_j(\mathbf{r}_a,\mathbf{r}', \tilde{\omega})=i\mu_0\sqrt{\frac{\epsilon_0
  \Im\epsilon(\mathbf{r'},\tilde{\omega})}{\pi}}\tilde{\omega}^2\sum_k(\mathbf{d}_a)_k\mathbf{G}_{kj}
  (\mathbf{r}_a,\mathbf{r}',\tilde{\omega})
\end{equation*}
We can formally solve Eqs. (\ref{eqbeta}) and (\ref{eqxi}) with vanishing initial values
of $\beta_j(\mathbf{r}',\tilde{\omega})$ and $\xi_{ab,j}(\mathbf{r}',\tilde{\omega})$.
The following relation will be used
in the calculation:
\begin{equation*}
\begin{aligned}
 \int_{j,\mathbf{r}'}g_j & (\mathbf{r}_a,\mathbf{r}',\tilde{\omega}) g^*_j(\mathbf{r}_b,\mathbf{r}',\tilde{\omega})\\
 & =\frac{\mu_0}{\pi}\tilde{\omega}^2\mathbf{d}_a\cdot\Im\mathbf{G}(\mathbf{r}_a,\mathbf{r}_b,\tilde{\omega})\cdot
\mathbf{d}_b \equiv  \Im g_{ab}(\tilde{\omega}).
\end{aligned}
\end{equation*}
Within the Markov approximation, we obtain the equation for the amplitudes of the individual emitters:
\begin{equation}\label{spalpha-1}
\begin{aligned}
  -\partial_t\alpha_a(t)= &
  \alpha_a(t)\int_{\tilde{\omega}}\Im g_{aa}(\tilde{\omega})\xi_t(\tilde{\omega},-\omega_{a,sg})\\
   +&\alpha_a(t)\sum_{b\neq a} \int_{\tilde{\omega}}\Im g_{bb}
   (\tilde{\omega})\xi_t(\tilde{\omega},\omega_{b,sg})  \\
   + & \sum_{b\neq a}\alpha_b(t)\int_{\tilde{\omega}}\Im g_{ab}(\tilde{\omega})
   \xi_t(\tilde{\omega}, -\omega_{b,sg})  \\
   + & \sum_{b\neq a}\alpha_b(t) \int_{\tilde{\omega}}\Im g_{ba}(\tilde{\omega})
   \xi_t(\tilde{\omega},\omega_{a,sg}).
\end{aligned}
\end{equation}
where
\begin{equation*}
  \xi_t(\omega_1, \omega_2)\equiv\frac{1-e^{-i(\omega_1+\omega_2)t}}{i(\omega_1+\omega_2)}.
\end{equation*}
The first and the third lines in Eq. (\ref{spalpha-1}) come from the ``rotating wave'' terms of the Hamiltonian,
while the second and the fourth lines are attributed to the ``counter-rotating wave'' terms.
Assuming $\omega_{i,sg}=\omega_{sg}$ and taking the long time limit
\begin{equation*}
\begin{aligned}
   \xi_t(\tilde{\omega},-\omega_{sg}) & \rightarrow
  \frac{-i}{\tilde{\omega}-\omega_{sg}-i\epsilon} \\
  \xi_t(\tilde{\omega},\omega_{sg}) & \rightarrow
  \frac{-i}{\tilde{\omega}+\omega_{sg}-i\epsilon},
\end{aligned}
\end{equation*}
Eq. (\ref{spalpha-1}) can be evaluated as
\begin{equation}\label{indi}
\begin{aligned}
  -\partial_t\alpha_a(t)= & \alpha_a\bigg[\pi \Im g_{aa}(\omega_{sg})
  -i\int_{\tilde{\omega}}\Im g_{aa}(\tilde{\omega})\mathcal{P}\frac{1}{\tilde{\omega}-\omega_{sg}}\bigg]\\
   & -i(N-1)\alpha_a\int_{\tilde{\omega}}\Im g_{bb}(\tilde{\omega})\mathcal{P}
\frac{1}{\tilde{\omega}+\omega_{sg}}\\
 + & \sum_{b\neq a}\alpha_b\bigg[ \pi \Im g_{ab}(\omega_{sg})-i\int_{\tilde{\omega}}
\Im g_{ab}(\tilde{\omega})\mathcal{P}\frac{1}{\tilde{\omega}-\omega_{sg}}\bigg],\\
& -i\sum_{b\neq a}\alpha_b\int_{\tilde{\omega}}\Im g_{ba}(\tilde{\omega})
\mathcal{P}\frac{1}{\tilde{\omega}+\omega_{sg}}.
\end{aligned}
\end{equation}
where $\mathcal{P}$ denotes the principal value integral and
we have assumed the translation symmetry, i.e., $\Im g_{bb}(\tilde{\omega})$ is
identical for all $b$.

Next, we shall use the Kramers-Kronig relation
\begin{equation*}
\begin{aligned}
  \Re g_{ab}(\omega_{sg}) & \equiv  \frac{\mu_0}{\pi}\tilde{\omega}^2\mathbf{d}_a\cdot\Re\mathbf{G}(\mathbf{r}_a,\mathbf{r}_b,\tilde{\omega})\cdot
\mathbf{d}_b \\
& = \frac{2}{\pi}\mathcal{P}\int_{0}^{\infty}d\tilde{\omega} \frac{\tilde{\omega}}{\tilde{\omega}^2-\omega_{sg}^2}\Im g_{ab}(\tilde{\omega}).
\end{aligned}
\end{equation*}
After organizing terms from all four lines, we have
\begin{equation*}
\begin{aligned}
  \partial_t\alpha_a(t)= & i\pi\sum_{b=1}^{N}\alpha_b(t)g_{ab}(\omega_{sg}) \\
  & +i(N-2)\alpha_a(t)\int_{\tilde{\omega}}\Im g_{aa}(\tilde{\omega})
\frac{1}{\tilde{\omega}+\omega_{sg}}.
\end{aligned}
\end{equation*}
We translate the amplitudes into wave vector representation
\begin{equation*}
  \alpha_{\mathbf{q}_{\shortparallel}}=\frac{1}{\sqrt{N}}\sum_{a=1}^{N}
  \alpha_a(t)e^{-i\mathbf{q}_{\shortparallel}\cdot\mathbf{r}_a}.
\end{equation*}
Then we obtain the equation for $\alpha_{\mathbf{q}_{\shortparallel}}$,
\begin{equation}
\begin{aligned}
  -i\partial_t\alpha_{\mathbf{q}_{\shortparallel}}(t) & =(N-2)\alpha_{\mathbf{q}_{\shortparallel}}(t)
  \int_{0}^{\infty}d\tilde{\omega}\,\frac{\Im g_{aa}(\tilde{\omega})}{\tilde{\omega}+\omega_{sg}}\\
  + \pi & N\int\frac{\mathbf{d^2k_{\shortparallel}}}{(2\pi)^2}\,\alpha_{\mathbf{k}_{\shortparallel}}(t)
  \zeta(\mathbf{q}_{\shortparallel},\mathbf{k}_{\shortparallel})
  g_{z_{at}}(\omega_{sg},\mathbf{k}_{\shortparallel}).
\end{aligned}
\end{equation}
Now we focus on $\mathbf{q}_{\shortparallel}=\mathbf{k}_{sp}$. Substituting the
approximation $\alpha_{\mathbf{k}_{\shortparallel}}=\alpha_{\mathbf{k}_{sp}}
\zeta(\mathbf{k}_{\shortparallel},\mathbf{k}_{sp})$ into the above equation yields
\begin{equation}\label{coll}
\begin{aligned}
  -i\partial_t\alpha_{\mathbf{k}_{sp}}(t) & =(N-2)\alpha_{\mathbf{k}_{sp}}(t)
  \int_{0}^{\infty}d\tilde{\omega}\,\frac{\Im g_{aa}(\tilde{\omega})}{\tilde{\omega}+\omega_{sg}}\\
  + \pi & N \alpha_{\mathbf{k}_{sp}}(t)\int\frac{\mathbf{d^2k_{\shortparallel}}}{(2\pi)^2}\,
  |\zeta(\mathbf{k}_{sp},\mathbf{k}_{\shortparallel})|^2
  g_{z_{at}}(\omega_{sg},\mathbf{k}_{\shortparallel}).
\end{aligned}
\end{equation}
The collective level shift $\delta_c \omega_{s}$ and decay rate $\gamma_{c}$ can be
extracted from the above equation,
\begin{subequations}
  \begin{align}
    \begin{split}\label{shifts}
  \delta_c\omega_s & = -(N-2)\int_{0}^{\infty}d\tilde{\omega}\,\frac{\Im g_{aa}(\tilde{\omega})}{\tilde{\omega}+\omega_{sg}} \\
       &\quad -\pi N\int\frac{\mathbf{d^2k_{\shortparallel}}}{(2\pi)^2}\,
  |\zeta(\mathbf{k}_{sp},\mathbf{k}_{\shortparallel})|^2
  \Re g_{z_{at}}(\omega_{sg},\mathbf{k}_{\shortparallel}),
\end{split}\\
\begin{split}\label{rates}
    \gamma_c &= \pi N\int\frac{\mathbf{d^2k_{\shortparallel}}}{(2\pi)^2}\,
  |\zeta(\mathbf{k}_{sp},\mathbf{k}_{\shortparallel})|^2
  \Im g_{z_{at}}(\omega_{sg},\mathbf{k}_{\shortparallel}).
  \end{split}
  \end{align}
\end{subequations}
Then we focus on the collective level shift of the emitter ground state $|G\rangle$.
We assume the ansatz
\begin{equation}
  |\Phi\rangle= \beta \ket{G,\varnothing}+
\int_{j,\tilde{\omega},\mathbf{r}'}\sum_{a=1}^{N} \alpha_{a,j}(\tilde{\omega},\mathbf{r}')
|e_a,1_{j,\tilde{\omega},\mathbf{r}'}\rangle\bigotimes_{b\neq a}|g_{b}\rangle,
\end{equation}
and let $\beta(t=0)=1$. With the Markov approximation, the equation of evolution for $\beta(t)$
gives the collective energy shift of the atomic ground state
\begin{equation}\label{shiftg}
  \delta_c\omega_g=- N \int_{\tilde{\omega}}\Im g_{aa}(\tilde{\omega})\frac{1}{\tilde{\omega}+\omega_{sg}}.
\end{equation}
The frequency shift between $|\psi_{\mathbf{k}_{sp}}\rangle$ and $|G\rangle$ is given as $\delta_c\omega_s-\delta_c\omega_g$.

For the collective Lamb shift of $\ket{\mathbf{k}_{sp}}$, we have to subtract the
the Lamb shift $\delta\omega_{sg}$ of the system with only a single emitter, which
can be obtained from Eqs. (\ref{indi}) and (\ref{shiftg}) with $N=1$:
\begin{equation}\label{indishift}
  \delta\omega_{sg}= -\int_{\tilde{\omega}}\Im g_{aa}(\tilde{\omega})\mathcal{P}\frac{2\omega_{sg}}{\tilde{\omega}^2-\omega_{sg}^2}.
\end{equation}

The collective Lamb shift of the single-SPP superradiance, $\Delta_c\omega_{sg}$, is hence
determined from Eqs. (\ref{shifts}), (\ref{shiftg}) and (\ref{indishift}) as
\begin{equation}
  \Delta_c\omega_{sg}=\delta_c\omega_s-\delta_c\omega_g-\delta\omega_{sg}.
\end{equation}

\providecommand{\latin}[1]{#1}
\makeatletter
\providecommand{\doi}
  {\begingroup\let\do\@makeother\dospecials
  \catcode`\{=1 \catcode`\}=2 \doi@aux}
\providecommand{\doi@aux}[1]{\endgroup\texttt{#1}}
\makeatother
\providecommand*\mcitethebibliography{\thebibliography}
\csname @ifundefined\endcsname{endmcitethebibliography}
  {\let\endmcitethebibliography\endthebibliography}{}


\end{document}